\documentclass[numbook]{svjour3}

\usepackage{booktabs}
\usepackage{graphicx}
\usepackage{natbib}
\usepackage{xcolor}

\usepackage{tcolorbox} 
\newtcolorbox{summary}{
    rounded corners,
    arc = 5pt
}

\usepackage{pifont}
\newcommand{\cmark}{\ding{51}}
\newcommand{\xmark}{\ding{55}}

\usepackage[hidelinks]{hyperref}
\usepackage{cleveref}

\newcommand{\google}{\textbf{\textit{google}}}
\newcommand{\popular}{\textbf{\textit{top1000}}}
\newcommand{\codeword}[1]{\texttt{\textbf{\small\color{black}#1}}}

\journalname{Empirical Software Engineering}

\begin{document}

\title{Detecting Outdated Code Element References in Software Repository Documentation
}

\author{Wen Siang Tan \and
        Markus Wagner \and
        Christoph Treude
}

\institute{Wen Siang Tan \at
               University of Adelaide \\
               \email{wensiang.tan@adelaide.edu.au}
           \and
           Markus Wagner \at
               University of Adelaide \\
               \email{markus.wagner@adelaide.edu.au}
           \and
           Christoph Treude \at
               University of Melbourne \\
               \email{christoph.treude@unimelb.edu.au}
}

\date{Received: date / Accepted: date}

\maketitle

\begin{abstract}
Outdated documentation is a pervasive problem in software development, preventing effective use of software, and misleading users and developers alike. We posit that one possible reason why documentation becomes out of sync so easily is that developers are unaware of when their source code modifications render the documentation obsolete. Ensuring that the documentation is always in sync with the source code takes considerable effort, especially for large codebases. To address this situation, we propose an approach that can automatically detect code element references that survive in the documentation after all source code instances have been deleted. In this work, we analysed over 3,000 GitHub projects and found that most projects contain at least one outdated code element reference at some point in their history. We submitted GitHub issues to real-world projects containing outdated references detected by our approach, some of which have already led to documentation fixes. As an initiative toward keeping documentation in software repositories up-to-date, we have made our implementation available for developers to scan their GitHub projects for outdated code element references.
\end{abstract}

\section{Introduction}
\label{sec:intro}

\sloppy

Outdated documentation is a common and well-known problem in software development~\citep{lee2019automatic}. It hinders the effectiveness of documentation~\citep{forward2002relevance}, prevents developers from using APIs and libraries efficiently~\citep{uddin2015api}, contributes to software ageing~\citep{parnas1994software} and confusion~\citep{kajko2005survey}, and it demotivates newcomers~\citep{steinmacher2018let}. In a recent study on software documentation issues, Ahgajani et al.~\citep{aghajani2019software} found that ``up-to-dateness problems'' account for 39\% of documentation content issues. Previous studies also revealed that more than two-thirds of participants surveyed believe that their system documentation is outdated~\citep{de2005study, lethbridge2003software}. Despite these findings, outdated documentation has remained an issue in the software engineering community due to the efforts needed to ensure that the documentation is in sync with the source code. Unlike source code, software documentation gets outdated ``silently'', i.e., there are no crashes or error messages to indicate that documentation is no longer up-to-date.\footnote{This is a well-known problem in software development, e.g., the documentation of tda-api states `TDA might change them at any time, at which point this document will become silently out of date', see \url{https://tda-api.readthedocs.io/en/latest/client.html}.} In many cases, developers are not aware that the source code changes they made have rendered the documentation outdated.

As a step toward helping developers to keep their documentation up-to-date, we propose an automated approach that detects outdated references in README file and wiki pages of a GitHub project. We focus our analysis on GitHub since it gives us access to the documentation of a large number of projects in a consistent format. We analysed the current state and full history of documentation of more than 3,000 GitHub projects and found that 28.9\% of the most popular projects on GitHub currently contain at least one outdated reference, with 82.3\% of the projects being outdated at least once during the project's history. These references were typically outdated for years before they were noticed and fixed by project maintainers.

The remainder of the paper is structured as follows: We motivate our work through a real-world example of outdated documentation in \Cref{sec:motivation}, explain our approach in \Cref{sec:approach}, and introduce the research questions in \Cref{sec:questions}. We report our findings in \Cref{sec:results}, present our publicly available implementation in \Cref{sec:implementation}, and interpret our findings in \Cref{sec:discussion}. We discuss the limitations of our approach in \Cref{sec:limitations} before we conclude the paper with related and future work in \Cref{sec:related,sec:conclusion}.

\section{Motivating Example}
\label{sec:motivation}

\begin{figure}[htbp]
    \centering
    \includegraphics[width=1.0\textwidth]{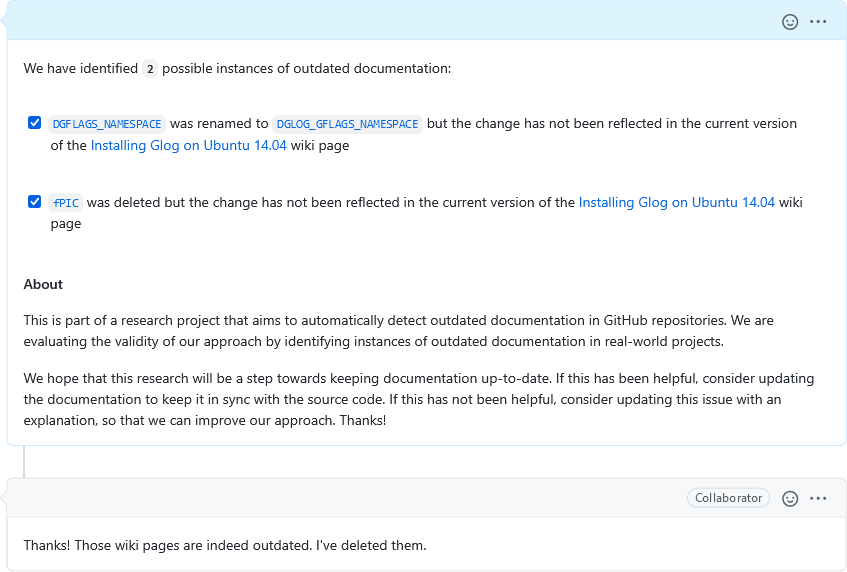}
    \caption{Screenshot of the GitHub issue submitted}
    \label{fig:glog}
\end{figure}

The google/glog project\footnote{\url{https://github.com/google/glog}} is one of the projects we found to contain outdated documentation. We detected an instance of the code element \codeword{DGFLAGS\_NAMESPACE} in the source code\footnote{\url{https://github.com/google/glog/blob/921651e97c3892e656287f1cfa923319f0799729/cmake/DetermineGflagsNamespace.cmake\#L36}} when the documentation was last updated. On 1 June 2018, the code element was renamed to \codeword{DGLOG\_GFLAGS\_NAMESPACE} in one of the commits.\footnote{\url{https://github.com/google/glog/commit/abce78806c8a93d99cf63a5a44ff09873f46b56f}} However, the documentation\footnote{\url{https://github.com/google/glog/wiki/Installing-Glog-on-Ubuntu-14.04/aa4fc07826bca7edf4aae57acd53119e515f9963}} was not updated to reflect the changes. In the same project, another code element \codeword{fPIC} was found 21 times in the source code\footnote{\url{https://github.com/google/glog/blob/921651e97c3892e656287f1cfa923319f0799729/m4/libtool.m4\#L3905}} when the documentation was last updated, but the document was not updated when all source code instances of the code element were deleted in this commit.\footnote{\url{https://github.com/google/glog/commit/b539557b3692c9c68d4e91d3cc920e8d14490d46}} We reported the discrepancies by submitting a GitHub issue\footnote{\url{https://github.com/google/glog/issues/750}} to the project's repository (\Cref{fig:glog}). Following our report, the project maintainer fixed the outdated documentation by deleting the document containing the two outdated references.

Much like this motivating example, source code and documentation often remain out of sync for some time before getting discovered. Our approach can automatically detect such discrepancies and enable project maintainers to monitor how source code and documentation evolve. The next section will discuss our approach in detail: (1) the criteria used to select documentation such as the README file and wiki pages in the project, (2) the method used to detect code elements such as \codeword{DGFLAGS\_NAMESPACE} and \codeword{fPIC} in the motivating example, (3) the steps needed to match code element references to actual instances in the source code, and (4) how the approach can be generalised to study the state of a project over time.

\section{Approach}
\label{sec:approach}

To detect outdated code element references in software repositories, relevant pieces of documentation need to be identified first. We extract from the documentation a list of potentially outdated references to code elements and match them to actual instances in the source code. If a reference remains in the documentation after all instances have been deleted from the source code, we consider the documentation outdated. The rest of this section describes this process in detail.

\subsection{Identifying documentation}
\label{sec:approach-sec1}

GitHub provides two main forms of documentation for project maintainers to document their projects. The README file is a convenient way to introduce the project to users and contributors. In a study by Prana et al.~\citep{prana2019categorizing} to categorise different types of content found in README files, the authors report that the majority of the README files from 393 randomly sampled projects contain some form of introduction or project background. In addition, README files often contain information for issues that may be encountered while using the project such as setup guides and API documentation. Project maintainers may also opt to make use of the wiki section for hosting documentation, which typically describes the project in more detail. One of the main differences between README and wiki is that the wiki may contain many pages while README is a single file. As any file types can be stored in GitHub wiki, only documentation written in file formats recognised by GitHub are considered in this work.\footnote{\url{https://github.com/github/markup}}

\begin{figure}[htbp]
    \centering
    \includegraphics[width=1.0\textwidth]{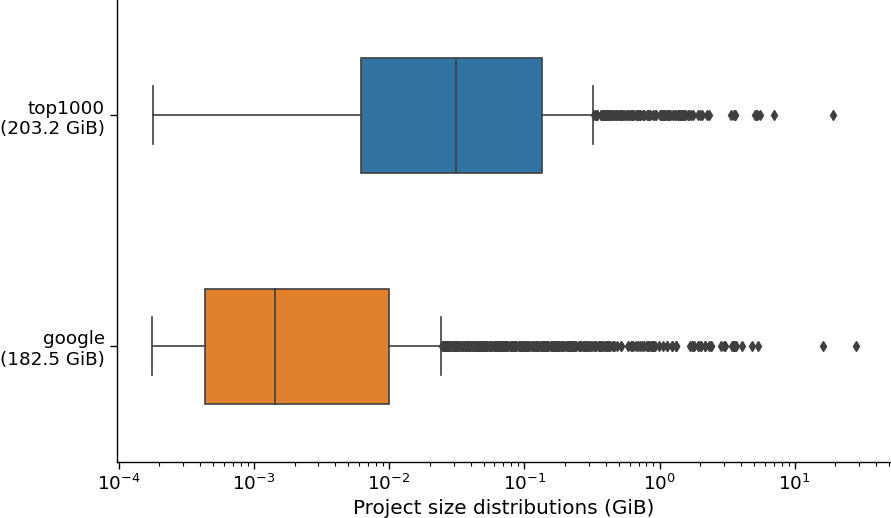}
    \caption{Project size distributions (GiB) for \popular{} and \google{} projects in log scale}
    \label{fig:project_size}
\end{figure}

\begin{figure}[htbp]
    \centering
    \includegraphics[width=1.0\textwidth]{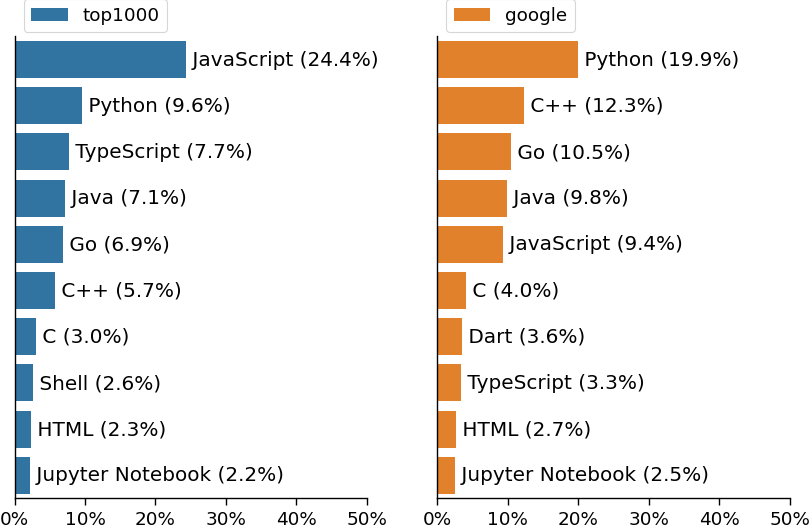}
    \caption{Top 10 programming languages used in \popular{} and \google{} projects}
    \label{fig:top_languages}
\end{figure}

We consider two datasets in this paper. The first dataset consists of the 1,000 most popular projects on GitHub, ranked by the number of stars.\footnote{\url{https://gitstar-ranking.com/repositories}, project names collected on 20 June 2022} The second dataset consists of all 2,279 GitHub projects from Google.\footnote{\url{https://github.com/orgs/google/repositories}, project names collected on 20 June 2022} \Cref{fig:project_size,fig:top_languages} show the size distributions and the top programming languages of \popular{} and \google{} projects. The list of project names for both datasets can be found in our online appendix.\footnote{\url{https://zenodo.org/record/7384588}}

\subsection{Extracting code elements}
\label{sec:approach-sec2}

In \Cref{sec:approach-sec1}, we identified a list of relevant documents from which we can extract potential outdated code element references. In this subsection, we outline the steps needed to extract such references from the documentation. These outdated references include variables, functions and class names found in the documentation. In this work, we use regular expressions to extract references to code elements in the documentation. Unlike parsers that are language-dependent, regular expressions can be used to extract possible candidates of outdated references in the documentation and matched to any source code files. We build on the work of Treude et al.~\citep{treude2014extracting} to extract code elements from the documentation using regular expressions, in which the authors have created a list of regular expressions to detect code elements.\footnote{\url{https://www.cs.mcgill.ca/~swevo/tasknavigator/}} As an example, one of the regular expressions \codeword{[A-Z][a-zA-Z]+ ?<[A-Z][a-zA-Z]*>} in that list is used to detect class templates such as the following:
\begin{itemize}
    \item \codeword{Worker<T>}
    \item \codeword{ArrayList<String>}
    \item \codeword{Callback<SimpleResponse>}
\end{itemize}

To help improve the quality of the list of code element references extracted from the documentation, i.e. code elements that are also found in the source code, we extracted a list of code elements using the original regular expression list and manually annotated if the reference is outdated. Each author annotated the same 50 randomly selected code elements\footnote{Previous work have used lesser than 50 data points to measure inter-rater agreement such as \url{https://link.springer.com/article/10.1007/s10664-021-10058-6}} detected from the \google{} projects to measure the inter-rater agreement. We achieved a free-marginal kappa of 0.92 when deciding whether the case is a true positive.

\begin{enumerate}
    \item We consider a code element reference as not outdated (false positive) if it fits any of the following criteria:
    \begin{enumerate}
        \item The source code file and documentation have identical content, e.g. one of the projects in our dataset contained their entire documentation corpus twice: once in the wiki and once as .md files in the source code repository.
        \item The code element reference extracted is a common word within the project (e.g. project name), a capitalised common word (PRIMARY, INACTIVE), an abbreviation (API, iOS), or a word that is not specific to the project (Data, User).
        \item The code element reference extracted from the documentation is a URL or URL alt text.
        \item The source code file is a text file that supposedly documents the project, e.g., an HTML file. 
        \item The code element matched in the source code is part of a source code comment.
    \end{enumerate}
    \item A reference is considered outdated (true positive) if the code element was found in a previous revision but has since been deleted:
    \begin{enumerate}
        \item The source code file exists in the current revision but the code element instance is deleted.
        \item The source code file is deleted in the current revision.
    \end{enumerate}
\end{enumerate}

During the manual annotation, we noticed that developers often use backticks (\`{}) in Markdown to indicate code elements. We also observed that extracting URLs from the documentation produced many code element references that are not matched to source code instances in a later stage. With the manual annotation data, we made a few modifications to the regular expression list:
\begin{enumerate}
    \item A regular expression to capture text enclosed in backticks is added. Code blocks (\`{}\`{}\`{}) are not added as they often contain longer texts that are less likely to be matched.
    \item A regular expression used to detect URLs in the original list is removed, URLs enclosed in backticks are still extracted.
    \item Many regular expression groupings in the original list are modified to extract only the code element, preventing additional spaces that are not part of the code element from getting extracted.
\end{enumerate}

The updated regular expression list used in this paper can be found in our online appendix.\footnote{\url{https://zenodo.org/record/7384588}}

\subsection{Matching code elements}
\label{sec:approach-sec3}

In the previous step, a list of potentially outdated references was extracted from the documentation using regular expressions. This subsection will describe the process of how these references are matched to actual instances in the source code to determine if they are outdated. In this work, a reference is considered outdated if the code element was found in both source code and documentation when the documentation was last updated, but the reference remains in the latest version of the documentation after all source code instances have been deleted (\Cref{tab:outdated}).
\begin{table}[htbp]
    \caption{What is outdated?}
    \begin{tabular}{@{}p{0.4\textwidth}*{2}{@{}p{0.3\textwidth}}@{}}
        \toprule
        & \textbf{Before} & \textbf{After} \\
        \midrule
        \textbf{Documentation} & \cmark & \cmark \\
        \textbf{Source code} & \cmark & \xmark \\
        \bottomrule
    \end{tabular}
    \label{tab:outdated}
\end{table}

To determine if a reference is currently outdated, we compare the number of instances found in two repository revisions. The first revision is the snapshot of the repository of when the documentation was last updated, and the second revision corresponds to the current revision of the repository. An instance is counted if it is a whole word, case-sensitive, and exact string match of the code element reference. If the number of source code instances goes from a positive integer (i.e. at least one code element instance was found in the source code when the documentation was updated) to a zero (i.e. all source code instances have been deleted in the current revision), we flag the reference as outdated. Going back to the motivating example, the two code element references flagged as outdated have the following number of instances found in the snapshot and the current repository revision (\Cref{tab:motivating_example}).
\begin{table}[htbp]
    \caption{Number of source code instances for the two code element references from the motivating example}
    \begin{tabular}{@{}p{0.4\textwidth}*{2}{@{}p{0.3\textwidth}}@{}}
        \toprule
        \textbf{Code element} & \textbf{Repository snapshot} & \textbf{Current revision} \\
        \midrule
        DGFLAGS\_NAMESPACE & 1 & 0 \\
        fPIC & 21 & 0 \\
        \bottomrule
    \end{tabular}
    \label{tab:motivating_example}
\end{table}

\begin{figure}[htbp]
    \centering
    \includegraphics[width=1.0\textwidth]{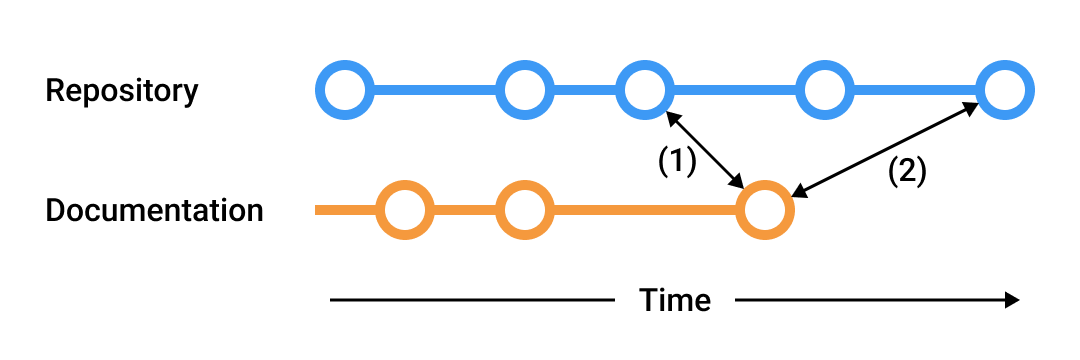}
    \caption{Linking the current documentation version to (1) repository snapshot and (2) current repository revision}
    \label{fig:link_normal}
\end{figure}

\paragraph{Linking references} On GitHub, a project's source code and wiki are stored separately in different Git repositories. We can get the snapshot of a project by interleaving the commit histories of both Git repositories: given a particular version of the documentation that is under investigation, we retrieve the most recent source code repository revision that was committed prior (\Cref{fig:link_normal}). In cases where the documentation is updated after the current repository revision, the snapshot refers to the current repository revision; this means that the number of instances found in both revisions are the same and the reference will not be flagged as outdated. This process is repeated for each code element reference extracted from the documentation to determine if the reference is currently outdated. Note that, as each page in the documentation may be updated at different times, code element references extracted from different pages may have a different repository snapshot.

\paragraph{File references} A code element reference may be incorrectly flagged as outdated when documentation references a file in the source code because file paths are often not explicitly written in the source code. To avoid flagging these cases as outdated, each variant of the file path that is an exact match of a code element reference is treated as an additional source code instance. In our implementation, a file path is considered a variant if it is a component of the file path including an optional slash at the beginning. For example, if the source code contains a file named \codeword{path/to/file.py}, all of the following variants are added to the list of code elements:
\begin{itemize}
    \item /path/to/file.py
    \item path/to/file.py
    \item /to/file.py
    \item to/file.py
    \item /file.py
    \item file.py
\end{itemize}

\subsection{Extending the analysis}
\label{sec:approach-sec4}

The approach outlined in the previous subsections can be generalised to analyse the state of documentation throughout a project's entire history. To help describe the state of a reference to code element C at the time of revision R and in document D, we designed a symbolic representation for the extended analysis:

\begin{itemize}
    \item \textbf{. (dot)} In revision R of the source code, document D did not exist.
    \item \textbf{- (dash)} In revision R of the source code, document D existed and it did not contain any references to C.
    \item \textbf{0} In revision R of the source code, document D existed and contained at least one reference to C and the source code did not contain any instances of C.
    \item \textbf{N} In revision R of the source code, document D existed and contained at least one reference to C and the source code contained an instance of C N times.
\end{itemize}

\begin{table}[htbp]
    \caption{Summary of symbolic representation used in the extended analysis}
    \begin{tabular}{@{}p{0.25\textwidth}*{3}{@{}p{0.25\textwidth}}@{}}
        \toprule
         & \textbf{Document existed\newline in revision R} & \textbf{Document has at\newline least one reference} & \textbf{Number of source\newline code instances} \\
        \midrule
        \textbf{. (dot)} & \xmark & & \\
        \textbf{- (dash)} & \cmark & \xmark & \\
        \textbf{0} & \cmark & \cmark & 0 \\
        \textbf{N} & \cmark & \cmark & N \\
        \bottomrule
    \end{tabular}
    \label{tab:symbolic_representation}
\end{table}

The symbolic representation can be summarised in \Cref{tab:symbolic_representation}. As an example, the first 50 revisions of the code element \codeword{renderFiles(`./files')} in the README file from the vuejs/vue-cli project\footnote{\url{https://github.com/vuejs/vue-cli}} have the following symbolic representation:

\begin{table}[htbp]
    \caption{Example of symbolic representation}
    \begin{tabular}{@{}p{1.0\textwidth}}
        \toprule
        . . . . . . . . . . . . . - - - - - - - - - - - - - - - - - - 3 3 3 3 3 3 3 0 0 0 0 - - - - - - - - \\
        \bottomrule
    \end{tabular}
    \label{tab:symbolic_example}
\end{table}

\begin{itemize}
    \item In the first 13 revisions, there is a dot (.) indicating that the README file did not yet exist.
    \item From revisions 14 to 31, there is a dash (-) indicating that the reference to the code element did not exist in the documentation (i.e., could not possibly be outdated).
    \item From revisions 32 to 38, there is a three (3) indicating that the reference to the code element existed in the documentation and was matched to three instances in the source code.
    \item From revisions 39 to 42, there is a zero (0) indicating that the reference to the code element existed in the documentation, but was no longer found in the source code (i.e., documentation was outdated).
    \item From revision 43 onward, there is a dash (-) again, indicating that the reference to the code element does not exist in the documentation anymore (i.e., documentation is no longer outdated).
\end{itemize}

\begin{figure}[htbp]
    \centering
    \includegraphics[width=1.0\textwidth]{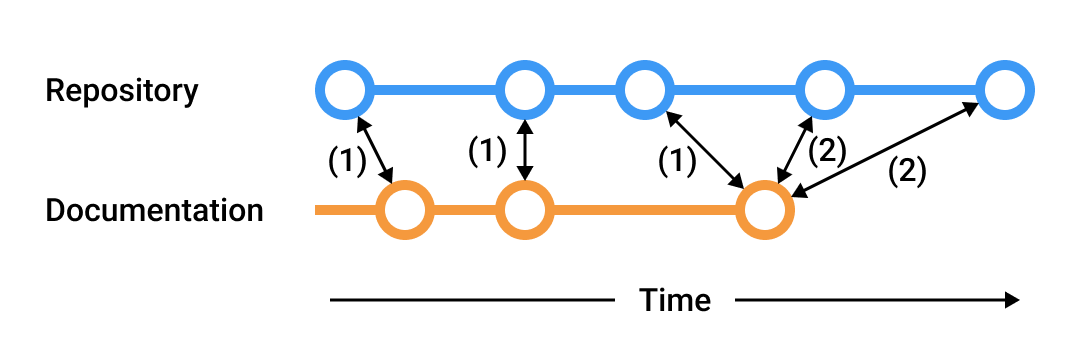}
    \caption{Linking each repository revision to a corresponding documentation version for repository commits made (1) before and (2) after the current documentation version}
    \label{fig:link_extended}
\end{figure}

\paragraph{Extending the linking process} To analyse the state of documentation throughout a project's history, we link each repository revision in the main branch to the next version of the documentation. Consistent with the method in \Cref{sec:approach-sec3}, the current version of the documentation is linked to the same repository revisions. \Cref{fig:link_extended} shows the links between repository revisions and their corresponding documentation versions.

\paragraph{Flagging as outdated} Consider a scenario where the symbolic representation of a particular code element in seven consecutive revisions is \codeword{2 0 0 . 0 0 0}. Two source code instances were found in the first revision and subsequently removed. The documentation was accidentally deleted in the fourth revision (indicated by the dot) and then restored (back to zero). Following the definition of outdated in \Cref{sec:approach-sec3} (positive integer followed immediately by a zero) will fail to flag this code element as outdated. Even though no source code instances are found in the latest revision, the reference still remains in the documentation. Using the symbolic representation, we can more accurately define `outdated' in the extended analysis. A code element is considered outdated if a positive integer is somewhere in front of a zero, even if it is not directly before the zero.

\paragraph{Creating a report} To make observing the trend of a code element throughout the project's history easier, we can record the number of code element instances found in each revision of the repository in a tabular form, grouped by their names and the documents from which they were extracted. \Cref{tab:report} shows a small section of the report from the vuejs/vue-cli project. We can see that three instances of the code element \codeword{renderFiles(`./files')} were found in revisions 37 and 38 followed by four zeros, which indicates that the code element reference was outdated from revisions 39 to 42. This was fixed in revision 43 when the outdated reference was deleted.
\begin{table}[htbp]
    \caption{A small section of the report generated from analysing the vuejs/vue-cli project (revision 37 to 43 for five code element references)}
    \begin{tabular}{@{}p{0.3\textwidth}*{7}{@{}p{0.1\textwidth}}@{}}
        \toprule
        \textbf{code element} & \textbf{R37} & \textbf{R38} & \textbf{R39} & \textbf{R40} & \textbf{R41} & \textbf{R42} & \textbf{R43} \\
        \midrule
        \textbf{projectOptions} & - & - & - & - & - & - & 7 \\
        \textbf{render(`./template')} & - & - & - & - & - & - & 3 \\
        \textbf{renderFiles(`./files')} & 3 & 3 & 0 & 0 & 0 & 0 & - \\
        \textbf{vue} & 198 & 205 & 205 & 205 & 205 & 210 & 210 \\
        \textbf{vue-cli-service} & 14 & 14 & 14 & 14 & 14 & 15 & 15 \\
        \bottomrule
    \end{tabular}
    \label{tab:report}
\end{table}

\section{Research Questions}
\label{sec:questions}

\noindent\textbf{RQ1: What is the current state of documentation?}
Our first research question investigates the current state of documentation in open-source projects on code element, document and project levels. This includes the number of code element references that are currently outdated and the duration for which they have been outdated.

\noindent\textbf{RQ2: What was the state of documentation during the projects' history?}
This research question aims to further explore the state of documentation by analysing the entire history of open-source projects. Similar to RQ1, we investigate the number of code element references that were outdated at some point in the project's history and the duration for which the outdated references typically survived in the documentation before getting fixed.

\noindent\textbf{RQ3: How do projects resolve their outdated documentation?}
After investigating the state of documentation in RQ1 and RQ2, we ask RQ3 to gain insights on how outdated documentation is typically fixed in real-world open-source projects by comparing the number of outdated references resolved by updating the source code, deleting the outdated code element reference, or by deleting the documentation.

\noindent\textbf{RQ4: How do open source projects respond to issues about outdated documentation?}
Our final research question examines how open-source project maintainers respond to our approach by creating GitHub issues highlighting the potentially outdated code element references detected in their projects.

\section{Results}
\label{sec:results}

This section will discuss the research questions raised in the previous section: (1) the current state of documentation, (2) the state of documentation over time, (3) how outdated documentation is commonly fixed, (4) and the responses of open source projects to our approach.

\begin{figure}[htbp]
    \centering
    \includegraphics[width=1.0\textwidth]{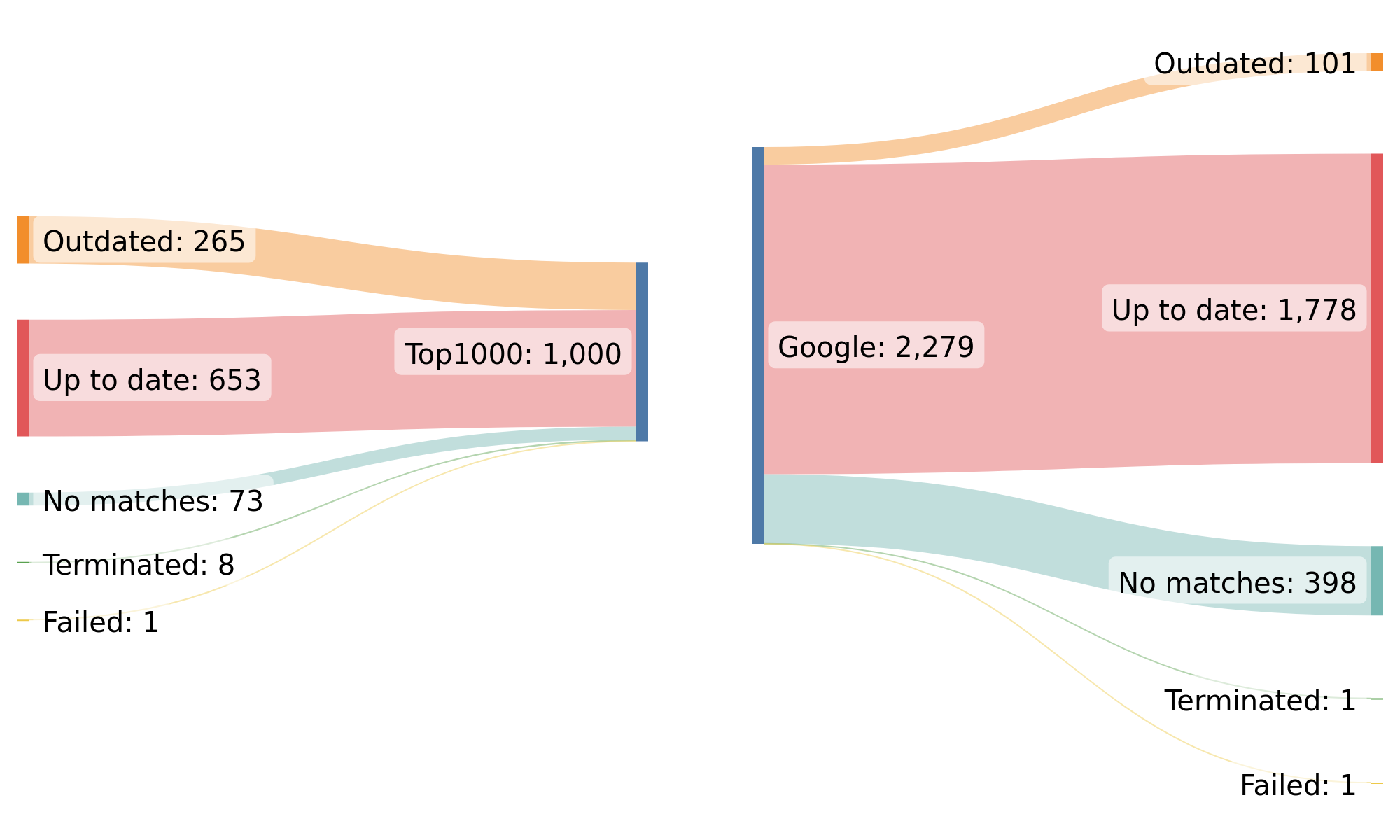}
    \caption{Analysis status of \popular{} and \google{} projects, indicating whether a repository's documentation is currently out of date}
    \label{fig:analysis_status}
\end{figure}

We ran our analysis on projects in the two datasets introduced in \Cref{sec:approach-sec1}. When cloning the repositories, one project\footnote{\url{https://github.com/google/material-design-icons}} failed due to a large number of files. In the \popular{} dataset, the analyses of 8 projects were terminated after failing to finish in a day. Among the 991 successfully analysed projects, 265 projects contained at least one outdated reference in their current version, 653 projects did not contain any outdated references and the documentation of 73 projects did not contain any matches to any code element in the source code. In addition, 90.4\% (896/991) of the \popular{} projects contained a README.md file and 60.0\% (595/991) had at least one wiki page at the time of analysis. In the \google{} dataset, the analysis of 1 project\footnote{\url{https://github.com/google/swiftshader}} was terminated after three days, leaving 2277 projects. The documentation of 101 projects was found to contain at least one outdated reference to a code element, the documentation of 1778 projects was up-to-date and the documentation of 398 projects did not contain code element references that were matched to the source code. 88.7\% (2019/2277) projects used a README.md file and 13.0\% (297/2277) used the wiki. \Cref{fig:analysis_status} shows the breakdown of the projects' statuses.

\subsection{RQ1: What is the current state of documentation?}

To investigate the current state of documentation in open-source projects, we scanned projects using the approach described in \Cref{sec:approach} and counted the number of projects for which the documentation contained at least one outdated code element reference (see \Cref{fig:analysis_status}). The same process is repeated at the document level to calculate the percentage of outdated documents. In addition, we can calculate the duration each code element reference is outdated for using the project's commit history.

\begin{figure}[htbp]
    \centering
    \includegraphics[width=1.0\textwidth]{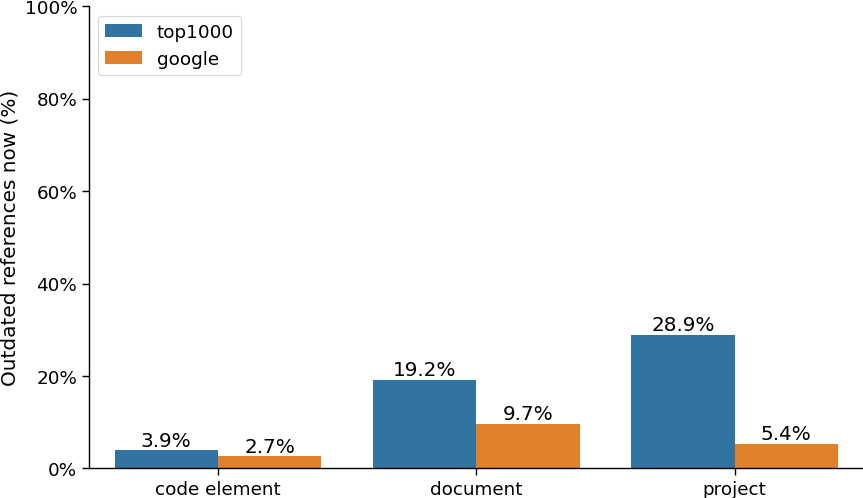}
    \caption{Percentage of references outdated at the time of analysis on code element, document and project levels}
    \label{fig:outdated_now}
\end{figure}

\begin{figure}[htbp]
    \centering
    \includegraphics[width=1.0\textwidth]{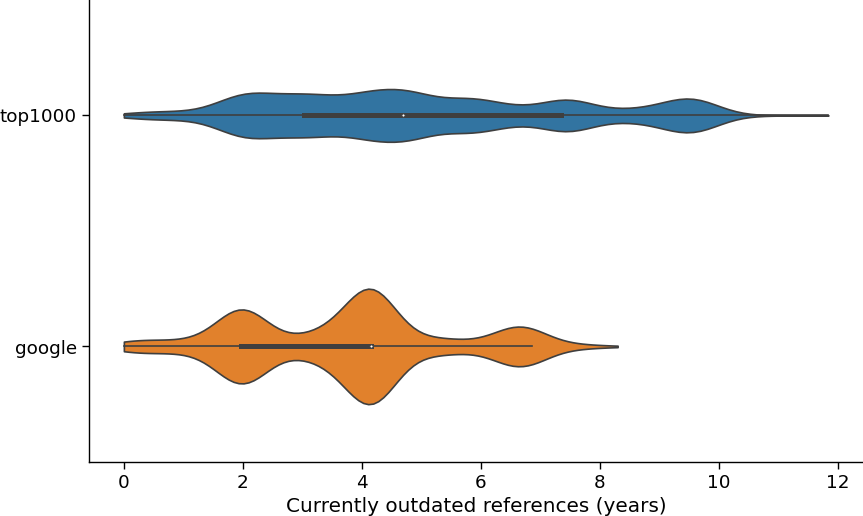}
    \caption{Distribution of duration that code element references have been outdated for at the time of analysis in \popular{} and \google{} projects}
    \label{fig:outdated_duration}
\end{figure}

In the \popular{} dataset, 3.9\% (7910/201852) of the code element references detected are currently outdated. We found that 19.2\% (1880/9784) of the documents contain at least one outdated reference to a code element, and 28.9\% (265/918) of the projects contain at least one outdated document. In the \google{} dataset, 2.7\% (1283/48078) code element references, 9.7\% (287/2947) documents, and 5.4\% (101/1879) projects are currently outdated (\Cref{fig:outdated_now}). On average, the references are currently outdated for 4.7 years for projects in the \popular{} dataset and 4.2 years for the \google{} dataset (\Cref{fig:outdated_duration}).

\begin{summary}
\textbf{RQ1 Summary} Documentation of 28.9\% \popular{} projects and 5.4\% \google{} projects were out of date at the time of analysis, with the references outdated for 4.7 and 4.2 years on average respectively.
\end{summary}

\subsection{RQ2: What was the state of documentation during the projects' history?}

\begin{figure}[htbp]
    \centering
    \includegraphics[width=1.0\textwidth]{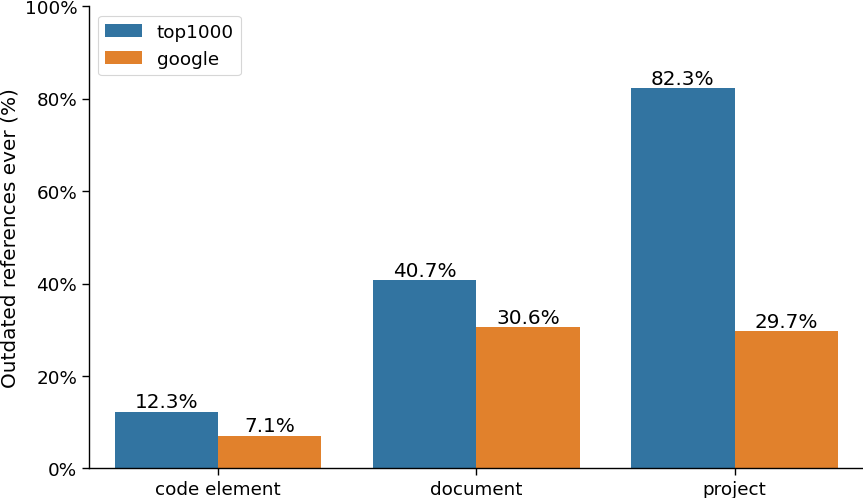}
    \caption{Percentage of references outdated at least once at some point during its history on code element, document and project levels}
    \label{fig:outdated_ever}
\end{figure}

\begin{figure}[htbp]
    \centering
    \includegraphics[width=1.0\textwidth]{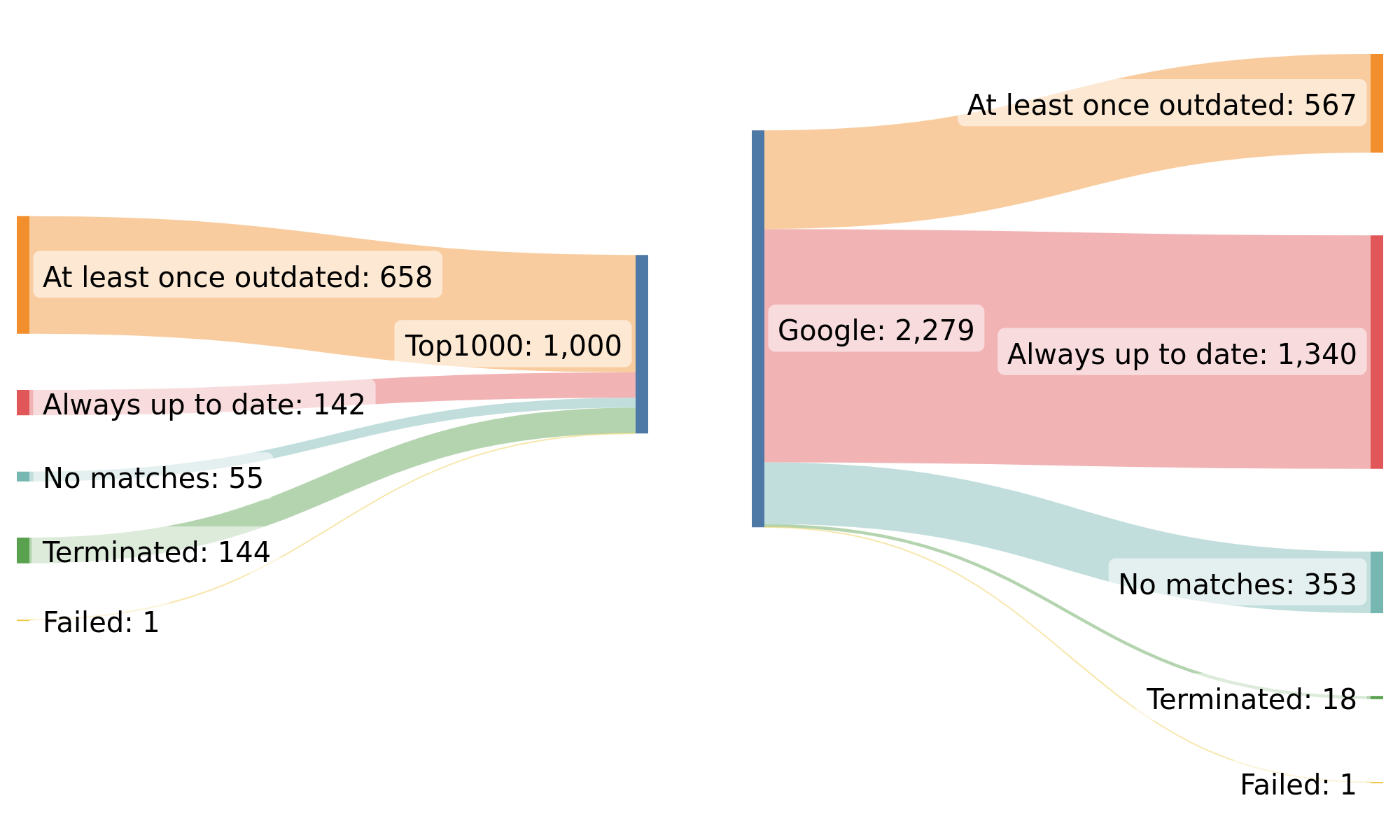}
    \caption{Extended analysis status of \popular{} and \google{} projects, indicating whether a repository's documentation was outdated at some point during its history}
    \label{fig:extended_analysis_status}
\end{figure}

To study how documentation evolves, we analysed the entire history of 800 projects from the \popular{} dataset. 82.3\% (658/800) of the projects, 40.7\% (2878/7071) of the documents, and 12.3\% (23588/191849) of the code element references are found to be outdated at some point in history. In addition, 1.3\% (2431/191849) of the code element references were outdated once again at some point in time after they were fixed. In addition, we analysed the full history of 1907 \google{} projects. 29.7\% (567/1907) projects, 30.6\% (925/3018) documents and 7.1\% (4176/58805) code elements were outdated sometime during the project's history (\Cref{fig:outdated_ever}). 0.4\% (210/58805) code element references were outdated again at least once after they were fixed. Note that the number of analysed projects for the extended analysis is different from the normal analysis (\Cref{fig:extended_analysis_status}).

\begin{figure}[htbp]
    \centering
    \includegraphics[width=1.0\textwidth]{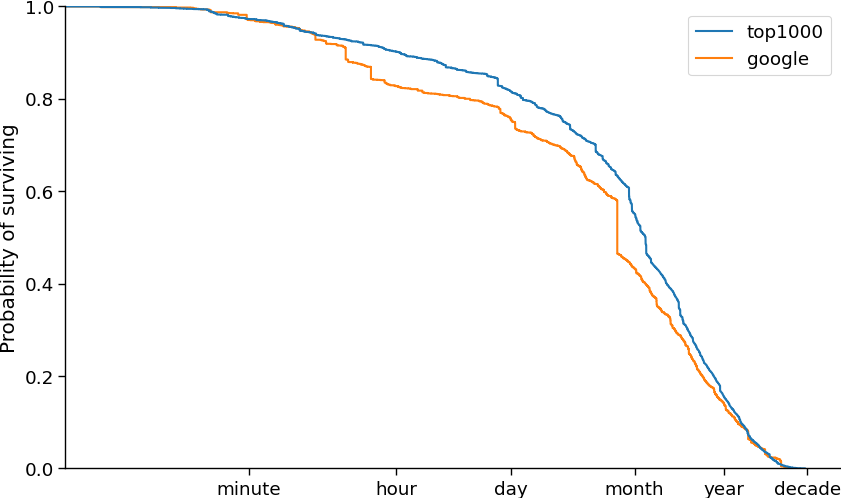}
    \caption{Time taken to fix outdated references in documentation for the \popular{} and \google{} dataset in log scale}
    \label{fig:outdated_curve}
\end{figure}

In addition to calculating the percentage of outdated documentation across project, document and code element levels, we calculated the duration of which outdated references survive in the documentation before getting fixed by project maintainers. \Cref{fig:outdated_curve} contains only outdated code elements references that project maintainers have already fixed with a timestamp difference greater than zero.\footnote{The babel/babel project had 7 negative timestamp differences caused by reverting README.md to an earlier version.} The probability of surviving is calculated by the percentage of outdated code element references that were still present in the documentation after the duration indicated by the x-axis has passed. For example, outdated references have around 55\% chance of surviving in \popular{} projects and 45\% in \google{} projects after a month.

\begin{summary}
\textbf{RQ2 Summary} Documentation of 82.3\% \popular{} projects and 29.7\% \google{} projects were outdated at some point in history, with 1.3\% and 0.4\% references outdated once again respectively after they were fixed.
\end{summary}

\subsection{RQ3: How do projects resolve their outdated documentation?}

There are three ways in which an outdated document can be resolved:
\begin{enumerate}
    \item Source code is changed to reintroduce code element instances, making the documentation in sync again.
    \item Documentation containing the outdated reference is updated to remove the outdated reference.
    \item Documentation containing the outdated reference is deleted, thereby removing the outdated reference.
\end{enumerate}

The three cases can be represented using the symbolic representation introduced in \Cref{sec:approach-sec4}:
\begin{table}[htbp]
    \caption{Types of documentation fixes}
        \begin{tabular}{@{}p{0.4\textwidth}*{2}{@{}p{0.3\textwidth}}@{}}
        \toprule
        & \textbf{Before} & \textbf{After} \\
        \midrule
        Documentation delete & 0 & . (dot) \\
        Documentation update & 0 & - (dash) \\
        Source code change & 0 & N \\
        \bottomrule
    \end{tabular}
    \label{tab:fixes}
\end{table}

Using the reports generated, we can study how the documentation was typically fixed throughout the project's history. For the \popular{} projects, we found that 73.6\% (17368/23588) outdated references to code elements were resolved throughout the projects' histories, with 47.6\% (8271/17368) fixed by changing the source code, 39.1\% (6783/17368) by updating the documentation, and 13.3\% (2314/17368) by deleting the documentation. For \google{} projects, 55.5\% (2319/4176) code element references were fixed by project maintainers. 50.2\% (1164/2319) were fixed by code changes, 43.3\% (1004/2319) by updating the documentation, and 6.5\% (151/2319) by deleting the documentation.

\begin{summary}
\textbf{RQ3 Summary} Project maintainers most commonly resolve outdated documentation by changing the source code, followed by updating and deleting the document to remove the outdated reference.
\end{summary}

\subsection{RQ4: How do open source projects respond to issues about outdated documentation?}

To examine the usefulness of our approach in real-world projects, we submitted GitHub issues to projects containing outdated references detected by our approach. In contrast to pull requests, creating an issue allows project maintainers to decide whether to delete the outdated reference in the documentation or update the documentation to reflect the changes made in the source code. Based on the manual annotation in \Cref{sec:approach-sec2}, we filtered projects from the \google{} dataset with at least one true positive and further narrowed them down to 15 actively maintained projects that have had new commits within the past year.

In the issues, we listed the outdated references with links to the documentation and an instance of the code element found in the source code. At the time of writing, 4 projects have responded positively, while the other 4 reported the issues as false positives. 7 projects have not yet responded to our GitHub issues. Across the 15 projects, we reported 19 instances of outdated documentation, 5 of which have been fixed by project maintainers. The following subsections will discuss two true positives and two false positives.

\begin{figure}[htbp]
    \centering
    \includegraphics[width=1.0\textwidth]{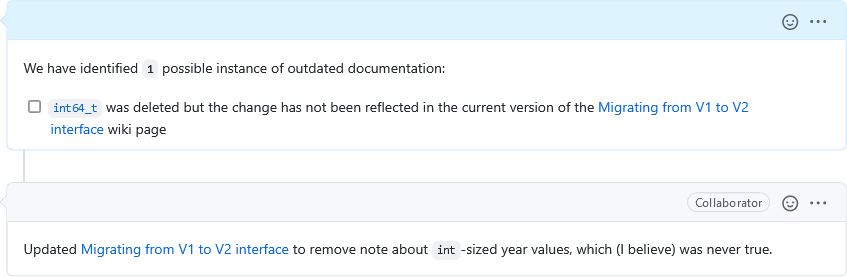}
    \caption{True positive: data type updated in the documentation}
    \label{fig:cctz}
\end{figure}

\begin{figure}[htbp]
    \centering
    \includegraphics[width=1.0\textwidth]{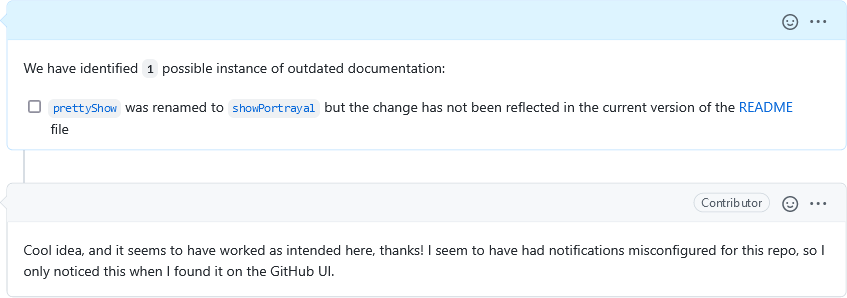}
    \caption{True positive: function name updated in the documentation}
    \label{fig:hs-portray}
\end{figure}

\paragraph{True positives} The cctz project was one of the projects that responded positively to our GitHub issue.\footnote{\url{https://github.com/google/cctz/issues/210}} In one of the commits, the code element instance \codeword{int64\_t} was removed entirely from the source code but the reference to the code element remained in the documentation. The project maintainer responded to our GitHub issue and updated the documentation to reflect the changes in the source code (\Cref{fig:cctz}). In the hs-portray project, the function \codeword{prettyShow} was renamed to \codeword{showPortrayal} in the source code, but the README file was not updated (\Cref{fig:hs-portray}). We alerted the developers of this discrepancy, and the issue was fixed subsequently.\footnote{\url{https://github.com/google/hs-portray/issues/7}}

\begin{figure}[htbp]
    \centering
    \includegraphics[width=1.0\textwidth]{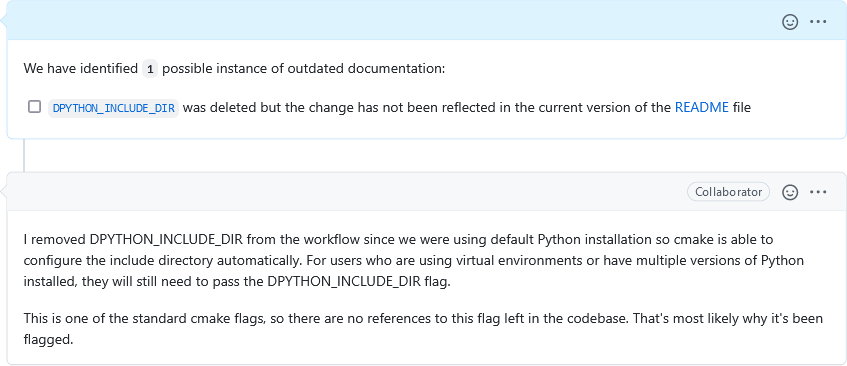}
    \caption{False positive: still relevant for users with multiple Python versions}
    \label{fig:clif}
\end{figure}

\begin{figure}[htbp]
    \centering
    \includegraphics[width=1.0\textwidth]{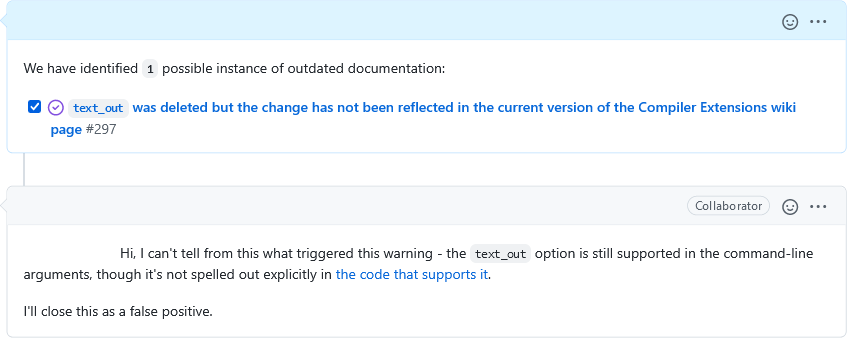}
    \caption{False positive: functionality remains in the program logic}
    \label{fig:gnostic}
\end{figure}

\paragraph{False positives} In one of the projects (\Cref{fig:clif}), a CMake flag was removed from the source code but the reference was not updated in the documentation. The project maintainers responded that the flag is no longer required in the source code but the documentation is still relevant for users that have installed multiple Python versions to configure the installation directory correctly.\footnote{\url{https://github.com/google/clif/issues/52}} A false positive was reported in another project (\Cref{fig:gnostic}) where the code element instance \codeword{text\_out} was deleted from the source code. Although the code element reference is not explicitly written in the source code, the functionality remains in the program logic which results in the code element reference getting falsely flagged as outdated.\footnote{\url{https://github.com/google/gnostic/issues/273}}

\begin{summary}
\textbf{RQ4 Summary} Several project maintainers responded positively to our GitHub issues and resolved the outdated references by updating or deleting the corresponding documents.
\end{summary}

\section{Implementation}
\label{sec:implementation}

The implementation of our approach called DOCER (Detecting Outdated Code Element References) is available in our online appendix.\footnote{\url{https://zenodo.org/record/7384588}} Running the script extracts code element references from the documentation and reports the number of code element instances found in the source code. The generated report includes additional information such as URLs to the source code, commit timestamps and SHAs to help developers investigate why a reference was flagged as outdated.

\section{Discussion}
\label{sec:discussion}

In this section, we will discuss our findings and the interesting differences between the two datasets used in this work. We investigated the current state of documentation in open-source software repositories and found that, on average, the \popular{} projects contain more outdated references than \google{} projects at the time of analysis. The references have also been outdated longer in the \popular{} projects (4.7 years) compared to \google{} projects (4.2 years). In the \popular{} dataset, 28.9\% of the projects were found to contain at least one outdated code element reference in contrast to 5.4\% of the \google{} projects. We hypothesise that this is because \google{} projects are generally smaller in size (median of 31.7 MiB for \popular{} projects and 1.47 MiB for \google{} projects), and hence easier for project maintainers to keep their documentation up-to-date.

In RQ2, we reviewed the full history of 800 \popular{} projects and 1907 \google{} projects. We found that 12.3\% and 7.1\% of the references to code elements detected respectively were outdated at some point in history, with the proportion higher on document and project levels. We investigated the sudden drops in survival probability for \google{} projects (\Cref{fig:outdated_curve}) and discovered that the biggest drop around the one month mark was caused by project maintainers deleting\footnote{\url{https://github.com/google/j2objc/commit/f9ff221f9eb8aacaecf057e3e9a1ca7c4e8a5beb}} and restoring\footnote{\url{https://github.com/google/j2objc/commit/592382e0bf314134fac9bfee862dacca50fccdb1}} large amount of source code files.

Next in RQ3, we looked into how open-source project maintainers usually resolve their outdated documentation. In our findings, approximately half of the fixes were attributed to source code changes. This is because the action of mass deleting and restoring source code files was interpreted as a fix caused by source code changes. We can also observe in various reports that the number of code element instances found in the source code suddenly drops to 0 and back to the original count.

Finally in RQ4, we examined the usefulness of our approach in real-world projects by alerting developers from 15 different Google projects of potential outdated references in their documentation where several project maintainers have responded positively to our GitHub issues. By using the implementation available in our online appendix, developers can scan for code element references that are potentially outdated in their GitHub project's documentation.

Although the content of this paper is centred around detecting outdated code element references in documentation hosted on GitHub projects, our approach can be generalised to other version control platforms. The next section of the paper will discuss the limitations of our approach.

\section{Limitations}
\label{sec:limitations}

In this work, our approach has identified many documents that are potentially outdated in software repositories but it does not detect all kinds of outdated documentation. As our approach relies on regular expressions for text extraction and matching, other forms of documentation containing outdated information such as images or videos cannot be detected. Our approach also cannot detect outdated relationships between the repository and documentation if the code elements are still present in the source code, i.e. documentation could be considered outdated even if all the code element references are matched.

Even though regular expressions allows us to easily extract code element references, they may sometimes lead to references being falsely categorised as outdated, e.g. deleting the final instance of a code element that is part of a source code comment. A project's change log may also be incorrectly flagged as outdated as it contains references to code elements that are no longer in the repository. These references should not be considered as outdated as they only serve as a notice for users that the referenced class or function has been deprecated. These false positives are difficult to eliminate and require project maintainers to verify individually.

\section{Related Work}
\label{sec:related}

In this section, we review related work on the impact of outdated documentation, efforts in the area of code element resolution, and work on detecting and/or fixing inconsistencies between source code and documentation. Our work is the first to detect outdated documentation based on references to code elements that are no longer in sync.

\subsection{Impact of outdated documentation}

According to the Open Source Survey~\citep{frances2017github}, ``incomplete or outdated documentation is a pervasive problem, observed by 93\% of respondents, yet 60\% of contributors say they rarely or never contribute to the documentation.'' In Sholler et al.'s `Ten simple rules for helping newcomers become contributors to open projects'~\citep{sholler2019ten}, the authors include ``Keep knowledge up-to-date and findable'' as one of their rules, arguing that ``outdated documentation may lead newcomers to a wrong understanding of the project, which is also demotivating. While it may be hard to keep material up-to-date, community members should at least remove or clearly mark outdated information. Signalling the absence or staleness of material can save newcomers time and also suggest opportunities for them to make contributions that they themselves would find useful.''

Outdated software documentation is a form of technical debt~\citep{kruchten2012technical} often referred to as documentation debt~\citep{aldaeej2021towards}. Rios et al.~\citep{rios2020hearing} list a number of effects of documentation debt, including low maintainability, delivery delay, rework, and low external quality, concluding that documentation debt affects several software development areas but especially requirements. With a similar focus on requirements, Mendes et al.~\citep{mendes2016impacts} report an extra maintenance effort caused by documentation debt of about 47\% of the total effort estimated for developing a project and an extra cost of about 48\% of the initial cost of the development phase. Compared to other types of technical debt, Liu et al.~\citep{liu2021exploratory} found that documentation debt is less commonly and more slowly removed.

Motivated by these findings, the goal of our work is the automated detection of outdated documentation, based on the intuition that documents can be considered outdated if they contain references to code elements that used to be part of a project but are no longer contained in a repository.

\subsection{Code element resolution}

Code element resolution refers to techniques that resolve a general (typically ambiguous) mention of a potential code element (e.g., a class or a method) to its definition~\citep{robillard2017demand}. Code element resolution has been employed in the context of emails~\citep{bacchelli2010linking}, tutorials~\citep{dagenais2012recovering}, or Stack Overflow~\citep{rigby2013discovering}, to name a few examples, often with the goal of linking relevant learning resources to code elements. Related work has also focused on automatically determining the importance of a code element mentioned in its context (e.g., in tutorial pages~\citep{petrosyan2015discovering}) or on detecting errors in API documentation~\citep{zhong2013detecting}.

Supervised machine learning approaches are often used for code element resolution, usually aiming at a balance of precision and recall. In this work, we rely on an improved version of the regular expressions used for code element detection by Treude et al.~\citep{treude2014extracting} and then use a very strict filter (exact match) to find instances of the mentioned code element in the source code. While this may underestimate the number of actually outdated code element references, we err on the side of caution to not establish traceability links that we are not confident about.

\subsection{Code-documentation inconsistencies}

Inconsistencies between source code and its documentation have been the target of various research efforts over the past years, with a particular focus on source code comments. Wen et al.~\citep{wen2019large} presented a large-scale empirical study of code-comment inconsistencies, revealing causes such as deprecation and refactoring. In one of the first attempts to detect and fix such inconsistencies, Tan et al.~\citep{tan2012tcomment} presented @tcomment for determining the correctness of Javadoc comments related to null values and exceptions. \textsc{DocRef} by Zhong and Su~\citep{zhong2013detecting} was designed to detect inconsistencies between source code and API documentation, based on the use of island parsing to extract code elements and reporting mismatched code elements as errors. AdDoc by Dagenais and Robillard~\citep{dagenais2014using} is a technique to identify code patterns in documentation using traceability links that can report new changes that do not conform to the code patterns of existing documentation. Also aimed at inconsistencies between source code and documentation, Ratol and Robillard~\citep{ratol2017detecting} presented Fraco, a tool to detect source code comments that are fragile with respect to identifier renaming.

Zhou et al.~\citep{zhou2018automatic} presented DRONE, a framework that can automatically detect defects in Java API documentation and generate meaningful natural language recommendations. This is achieved through a combination of static program analysis, part-of-speech tagging, and constraint solving. Another related work is FreshDoc, which is an approach proposed by Lee et al.~\citep{lee2019automatic} to automatically update class, method, and field names in the API documentation. This is done by extracting code elements with a grammar parser and analysing different versions of the source code. More recently, Panthaplackel et al.~\citep{panthaplackel2020learning} proposed an approach to automatically update existing comments when the source code is modified. This is accomplished by tokenising the comments and source code, and then modifying the comment tokens associated with the changes in source code.

In contrast to these related work, our approach detects outdated references to code elements in the documentation. To the best of our knowledge, there are currently no similar contributions for automatically detecting outdated documentation in software repositories when source code and documentation go out of sync.

\section{Conclusion/Future work}
\label{sec:conclusion}

In this paper, we proposed an approach that can automatically detect outdated references to code elements caused by removing all source code instances. We investigated the current state of documentation in software repositories, extended the approach to analyse the state of documentation throughout projects' history, explored how outdated documentation is resolved in open source projects, and with the results, we alerted Google developers of potentially outdated code element references in their projects.

In detail, we found that the majority of the most popular projects on GitHub contained at least one outdated reference to a code element at some point during their history and these outdated references usually survived in the documentation for years before they were fixed. By analysing the full history of projects, we discovered that outdated references are more likely fixed by updating the source code or document than deleting the entire document. Moreover, our GitHub issues have led to instances of outdated documentation getting fixed in real-world projects.

Although documentation gets outdated without warnings, developers can take steps to keep their documentation up-to-date by checking if the documentation needs to be updated whenever changes are made to the source code. Using our current implementation, developers have to manually scan their repository with each commit which may be repetitive and time consuming. A possible direction for future work is to create a workflow that automatically clones the repository, runs the analysis, and outputs the potentially outdated references. Using a tool such as GitHub Action\footnote{\url{https://github.com/features/actions}} to automate the workflow simplifies the process considerably as it allows developers to configure their repository to automatically scan for outdated references whenever there is a new commit or pull request.

Another potential direction for future work is expanding our approach to other forms of documentation, such as images. We hypothesise that texts in images are more likely to be outdated as they generally require more effort to update. These texts may be extracted using methods such as Optical Character Recognition, allowing us to detect more potentially outdated references. Applying customised sets of regular expressions for files written in different programming languages may also be another direction to help with more accurate matches in the source code, e.g. avoiding matching source code comments. We hope that this research will be a step toward keeping documentation in software repositories up-to-date.


%
%

\end{document}